\documentclass{article}    
\usepackage{graphicx}          
\usepackage{amsmath}
\usepackage{amsfonts}
\usepackage{amssymb}
\usepackage{xcolor}
\usepackage{algorithm}
\usepackage{algpseudocode}

\def\WW{\mathcal{W}}
\def\ZZ{\mathcal{Z}}
\def\XX{\mathcal{X}}
\def\UU{\mathcal{U}}
\def\VV{\mathcal{V}}
\def\SSS{\mathcal{S}}
\def\QQ{\mathcal{Q}}
\def\KK{{K}}
\def\xx{{x}}
\def\uu{{u}}
\def\vv{{v}}
\def\zz{{z}}
\def\rr{{r}}
\def\AA{{A}}
\def\BB{{B}}
\def\ee{{e}}
\def\ww{{w}}

\graphicspath{{figs/}}
\def\NoNumber#1{{\def\alglinenumber##1{}\State #1}\addtocounter{ALG@line}{-1}}

\newtheorem{assum}{Assumption}
\newtheorem{thm}{Theorem}
\newtheorem{proof}{Proof}
\newtheorem{rem}{Remark}
\newtheorem{cor}{Corollary}

\title{Robust Variable-Horizon MPC\\ with Adaptive Terminal Constraints}
\author{Renato Quartullo, Gianni Bianchini, Andrea Garulli, Antonio Giannitrapani}
\date{University of Siena, 53100, Siena, Italy}

\begin{document}
	
	\maketitle
	
	\begin{abstract}                          
		This paper presents a novel robust variable-horizon model predictive control scheme designed to intercept a  target moving along a known trajectory, in finite time. Linear discrete-time systems affected by bounded process disturbances	are considered and a tube-based MPC approach is adopted. The main contribution is an adaptive mechanism for choosing the terminal constraint set sequence in the MPC optimization problem. This mechanism is designed to ensure recursive feasibility while promoting minimization of the final distance to the target. Finite-time convergence of the proposed control scheme is proven. In order to evaluate its effectiveness, the designed control law is tested through numerical simulations, including a case study involving orbital rendezvous of a satellite with a tumbling object. The results indicate a significant reduction in conservatism compared to existing state-of-the-art methods using a fixed terminal set sequence.
	\end{abstract}
	\section{Introduction}
	Model Predictive Control (MPC) techniques usually require, at each time instant, the solution of an optimal control problem whose cost is a suitable function of the state and input variables, over a fixed-length prediction horizon \cite{rawlings2017model}.
	In recent years, there has been a growing interest towards MPC settings in which the length of the prediction horizon is itself a variable of the optimization problem. 
	In fact, in some applications it is crucial to perform the desired task in the least amount of time or within a prescribed deadline. This is the case, for example, of spacecraft rendezvous and docking maneuvers with moving targets \cite{boyarko2011optimal,di2012model,capello2018flyable,leomanni2022variable,ramirez2023collision}, and  motion planning of robotic systems \cite{verscheure2009time,nascimento2018nonholonomic,ardakani2018model}. 
	In such cases, it may indeed be useful to include the length of the prediction horizon in the cost function of the MPC scheme.
	
	A first possibility is to directly minimize the length of the prediction horizon. This results in the so-called \emph{minimum-time} or \emph{time-optimal MPC}.
	In \cite{van2011model}, a bilevel optimization scheme is proposed. The inner problem is a standard optimal control problem with fixed horizon length, while the outer problem aims at minimizing the horizon length under the constraint that the inner problem is feasible. 
	In \cite{rosmann2015timed}, time optimality is pursued by using time-elastic bands, i.e, by including the sampling time among the optimization variables.
	Time-optimal MPC for point-to-point motion is addressed in \cite{verschueren2017stabilizing}, where an exponential increase of the stage costs along the horizon is used to enforce time-optimality and closed-loop stability of the resulting control scheme. Lyapunov stability of minimum-time MPC has been studied in \cite{krener2018adaptive,sutherland2019closed}.
	
	An alternative to time-optimal MPC is known as \emph{shrinking-time MPC}: in this case, the horizon length is not an optimization variable, but it is reduced by design at every time step. This approach has been employed for an helicopter landing maneuver in \cite{greer2020shrinking}, and for energy-efficient operation of trains in  \cite{farooqi2020shrinking}. It is worth observing that the latter work is the only one among those cited so far in which some source of uncertainty is considered. In particular, suitable relaxations are proposed to retain recursive feasibility in the presence of bounded input disturbances.
	
	A more general setting is that of \emph{Variable-Horizon MPC (VH-MPC)}, in which the prediction horizon length is treated as a variable of the optimization problem and weighted in the cost function. In the seminal paper \cite{richards2006robust}, this framework is considered for linear systems affected by bounded process disturbances, with application to vehicle maneuvers. Recursive feasibility and finite-time convergence to a suitable region containing the target is achieved via a tube-based MPC approach.
	A possible drawback of this solution is that the convergence set is determined by the \emph{a priori} worst-case disturbance sequence, along the entire maneuver. Therefore, the resulting performance may be conservative in terms of final distance to the target.
	The VH-MPC approach has been recently applied to path tracking for autonomous vehicles \cite{wang2021path}, helicopter landing \cite{ngo2022variable}, satellite docking to a tumbling target~\cite{leomanni2022variable}, drone rendezvous with a moving platform \cite{persson2024optimization}.

	In this paper, a new robust VH-MPC scheme is proposed, whose aim is to intercept a moving target with known trajectory, in the presence of bounded process disturbance.
	The main novelty is an adaptation mechanism for the terminal set of the tube-MPC optimization problem, which has the twofold objective of preserving recursive feasibility while reducing as much as possible the final distance to the target. Finite-time convergence of the designed control scheme is proven. Moreover, as a side contribution, it is shown that time-optimal MPC can be recovered as a special case of the proposed VH-MPC setting, in which the resulting prediction horizon is guaranteed to reduce at every time step. 
	The VH-MPC scheme is tested on a simple example, to illustrate its main properties, and on a realistic case study featuring the rendezvous of a satellite with a tumbling target.
	Simulation results show that online adaptation of the terminal set leads to a remarkable reduction of conservatism, with respect to employing the terminal set corresponding to the a priori worst-case disturbance sequence.
	
	The rest of the paper is organized as follows. The variable-horizon problem is formulated as a tube-MPC scheme in Section~\ref{sec:PF}. The novel control strategy guaranteeing finite-time convergence is presented in Section~\ref{sec:algo} and then tested in the aforementioned simulation scenarios in Section~\ref{sec:simulations}. Some conclusions are drawn in Section~\ref{sec:conclusions}.
	
	\subsection*{Notation}
	The adopted notation is fairly standard. The symbols $\oplus$ and $\ominus$ represent the Minkowski sum and Pointryagin set difference, respectively. In particular, given two sets $\mathcal{A}$, $\mathcal{B}$, $\mathcal{A}\oplus\mathcal{B} = \{a+b\,:\,a\in\mathcal{A},b\in\mathcal{B}\}$ and $\mathcal{A}\ominus\mathcal{B} = \{a\,: \{a\}\oplus\mathcal{B}\subseteq\mathcal{A}\}$. The notation $\lfloor c \rfloor$ is used to indicate the integer part of $c\in\mathbb{R}$.
	
	\section{Problem Formulation}\label{sec:PF}
	Consider the linear time-invariant system
	\begin{equation}\label{eq:sys}
		\xx(k+1) = \AA\xx(k) + \BB\uu(k) + \ww(k),
	\end{equation}
	where $\xx(k) \in\mathbb{R}^n$ is the system state at time $k$, $\uu(k) \in\mathbb{R}^m$ is the control input and $\ww(k) \in \WW$ is a bounded disturbance. The control objective is to drive the state to a predefined reference trajectory $\rr(k)$ in finite time, while optimizing a suitable performance index. Moreover, the state and input are subject to the constraints
	\begin{equation}\label{eq:constraint}
		\xx(k) \in \XX(k),\,\uu(k) \in \UU(k),
	\end{equation}
	being $\XX(k),\,\UU(k)$ and $\WW$ convex sets.
	
	Consider now the disturbance-free (nominal) system, given by 
	\begin{equation}\label{eq:sys_nominal}
		\zz(k+1) = \AA\zz(k) + \BB\vv(k),
	\end{equation}
	and define the error between the true and the nominal state as $\ee(k) = \xx(k)-\zz(k)$. As in standard tube-based MPC \cite{rawlings2017model,chisci2001systems}, let the feedback policy be designed as
	\begin{equation}\label{eq:u_RMPC}
		\uu(k) = \vv(k) + \KK(\xx(k) - \zz(k)),
	\end{equation} 
	where $\vv(k)$ is the nominal control input and $\KK$ is such that $\AA_{\KK} = \AA+\BB\KK$ is Schur.	Then, the error system evolves as
	\begin{equation}\label{eq:kolma_gilbert}
		\ee(k+1) = \AA_{\KK}\ee(k) + \ww(k).
	\end{equation}
	It is well-known that if $\ee(0) = 0$ (i.e., $\zz(0)=\xx(0)$), then $\ee(k)\in \SSS(k),\, \forall k\geq 0$,  with
	\begin{equation}\label{eq:Sk}
		\SSS(k) = \begin{cases}
			\{0\}, & \text{if } k=0\\
			\sum_{i=0}^{k-1} \AA_{\KK}^i\WW, & \text{if } k\geq 1.
		\end{cases}
	\end{equation}
	Moreover, if the nominal state and input satisfy the tightened constraints,
	\begin{equation}\label{eq:constraint_tight}
		\begin{array}{l}
			\zz(k)\in\ZZ(k) = \XX(k) \ominus \SSS(k), \\
			\vv(k)\in \VV(k) = \UU(k)\ominus\KK\SSS(k),
		\end{array}
	\end{equation}
	then the state-input constraints~\eqref{eq:constraint} are robustly satisfied with respect to any disturbance realization $\ww(k)$.
	
	In this work, the performance index to be optimized weights the command input and state deviation from the reference trajectory. Moreover, in order to promote finite-time convergence it includes the horizon length $N$, thus resulting in the cost function
	\begin{equation}\label{eq:cost}
		J = N + \gamma_z \sum_{k=0}^{N} \|\zz(k)-\rr(k)\| + \gamma_v \sum_{k=0}^{N-1} \|\vv(k)\|,
	\end{equation}
	where $\gamma_z,\,\gamma_v \geq 0$ and $\|\cdot\|$ indicates any (possibly weighted) vector norm.
	
	The nominal control input $\vv(k)$ in \eqref{eq:u_RMPC} is computed via a robust variable-horizon MPC scheme. In particular, at each time instant $k$, the following optimization problem $\mathbb{P}_k(\xx(k),\ZZ_{f,k})$ is considered:
	\begin{equation}\label{eq:mpc_tube}
		\begin{aligned}
			\underset{N_k,\boldsymbol{\vv}_k,\boldsymbol{\zz}_k}{\text{min}} &  J_k=   N_k +  \gamma_z  \sum_{j=0}^{N_k}  \|\zz_k(j) - \rr(k + j)\| +  \gamma_v  \sum_{j=0}^{N_k-1}  \|\vv_k(j)\|\\[1mm]
			\text{s.t.} \quad & {\zz_k}(0)= \xx(k)\\[1mm]
			& {\zz_k}(j+1)=\AA {\zz_k}(j)+\BB{\vv_k}(j)  \\[1mm]
			&  {\zz_k}(j)\in \ZZ_k(j), \quad j=1,\ldots,N_k-1\\[1mm]
			&  \vv_k(j)\in \VV_k(j), \quad j=0,\ldots, N_k-1\\[1mm]
			&  {\zz_k}(N_k)\in\{\rr(k+N_k)\}\oplus\ZZ_{f,k}\\[1mm]
			&  N_k\in\mathbb{N^+},
		\end{aligned}
	\end{equation}
	where the sequences $\boldsymbol{\vv}_k= \left[\vv_k(0),\ldots,\,\vv_k(N_k-1)\right]$ and $\boldsymbol{\zz}_k = \left[\zz_k(0),\ldots,\,\zz_k(N_k)\right]$ are the nominal inputs and states, respectively, along the prediction horizon, and
	\begin{equation}\label{eq:constraint_tight_k}
		\begin{array}{l}
			\ZZ_k(j) = \XX(k+j)\ominus \SSS(j),\\ 
			\VV_k(j) = \UU(k+j)\ominus \KK\SSS(j)
		\end{array}
	\end{equation}
	are the tightened state and input constraint sets. 	
	In problem~\eqref{eq:mpc_tube}, $\ZZ_{f,k}$ is a sequence of convex terminal constraint sets to be properly designed.
	The solution of $\mathbb{P}_k(\xx(k),\ZZ_{f,k})$ consists of the optimal cost $J_k^*$, the optimal horizon length $N_k^*$, and the optimal input and state sequences, $\boldsymbol{\vv}_k^*$ and $\boldsymbol{\zz}_k^*$, respectively. At each step $k$, the first sample of the optimal control sequence is applied to system~\eqref{eq:sys}, i.e. $\uu(k) = \vv_k^*(0)$, as in the typical receding-horizon approach. From standard tube-MPC arguments, it turns out that, if \eqref{eq:mpc_tube} admits a solution satisfying the tightened constraints in~\eqref{eq:constraint_tight_k}, then the state and control trajectories of the uncertain closed-loop system \eqref{eq:sys}-\eqref{eq:u_RMPC} robustly satisfy the original constraints~\eqref{eq:constraint}.
	
	In order to ensure recursive feasibility and finite-time convergence of the robust VH-MPC scheme described above, a possible solution is the one proposed in~\cite{richards2006robust}. That study considers a specific instance of problem $\mathbb{P}_k(\xx(k),\ZZ_{f,k})$, in which $\rr(k) = 0,\,\forall k$ and $\ZZ_{f,k}$ is selected as 
	\begin{equation}\label{eq:Zf_How}
		\ZZ_{f,k} = \QQ \ominus \SSS(N_k),
	\end{equation} 
	where $\SSS(N_k)$ is defined as in~\eqref{eq:Sk} and $\QQ$ is a suitable set guaranteeing $\ZZ_{f,k}$ to be non-empty for each value of $N_k$. A reasonable choice for this set is $\QQ = \SSS(\infty)=\lim_{k\rightarrow \infty} \SSS(k)$. The set $\SSS(\infty)$ is the largest robustly positive invariant set for the error system~\eqref{eq:kolma_gilbert} and it can be computed as suggested, e.g., in~\cite{kolmanovsky1998theory}. The terminal set sequence~\eqref{eq:Zf_How} guarantees recursive feasibility of problem~\eqref{eq:mpc_tube} and finite-time convergence of the error to the set $\SSS(\infty)$, provided that $\gamma_v$ and $\gamma_z$ are suitably selected (see~\cite{richards2006robust} for details). In turn, this often results in a large terminal distance from the target at maneuver completion, thus leading to inaccurate target interception.
	
	
	In the next section, an adaptive way of choosing the terminal constraint sets $\ZZ_{f,k}$ is proposed.
	
	\section{VH-MPC with Adaptive Terminal Constraints}\label{sec:algo}
	In the perspective of intercepting a reference trajectory $\rr(k)$, the ideal choice for the terminal set would be $\ZZ_{f,k}=\{0\},\,\forall k$, corresponding to the equality terminal constraint
	\begin{equation}\label{eq:terminal_equality}
		\zz_k(N_k)=\rr(k+N_k).
	\end{equation}
	However, due to the presence of disturbances, such constraint may render the control problem~\eqref{eq:mpc_tube} infeasible. To this purpose, in this section, we propose a control strategy that relies on an adaptive choice of the terminal set $\ZZ_{f,k}$, in order to reduce the final distance to the desired trajectory, while guaranteeing both recursive feasibility and finite-time convergence of the control scheme.
	
	The idea is to keep solving problem~\eqref{eq:mpc_tube} with terminal equality constraint~\eqref{eq:terminal_equality}, namely $\mathbb{P}_k(\xx(k),\{0\})$, as long as the problem is feasible and the optimal cost decreases by at least a suitable positive quantity $\lambda$ with respect to the previous step.
	Whenever this does not occur, problem \eqref{eq:mpc_tube} is solved with a suitable terminal set $\ZZ_{f,k}$ that guarantees a cost decrease not smaller than $\lambda$. More specifically, the proposed control algorithm consists of the following procedure. At each time $k$, solve problem  $\mathbb{P}_k(\xx(k),\{0\})$ and let $\tilde{J}_k^*$ be the resulting optimal cost ($\tilde{J}^*_k=\infty$ if the problem is infeasible). Then:
	\begin{itemize}
		\item if $\tilde{J}_k^* \leq J^*_{k-1}-\lambda$, use the solution of $\mathbb{P}_k(\xx(k),\{0\})$ at time $k$, set $\ZZ_{f,k}=\{0\}$ and proceed to the next time step;
		\item if $\tilde{J}_k^* > J^*_{k-1}-\lambda$, solve $\mathbb{P}_k(\xx(k),\ZZ_{f,k})$, where 
		\begin{equation}\label{eq:Zfk}
			\ZZ_{f,k} = \ZZ_{f,k-1} \oplus \AA_K^{N^*_{k-1}-1}\WW,
		\end{equation}
		with the additional constraint $N_k \leq N^*_{k-1}-1$.
		Use the solution of such a problem
		as optimal at time $k$ and proceed to the next time step.
	\end{itemize}
	The overall control strategy is detailed in Algorithm~\ref{alg:control}. 
	\begin{algorithm}[h]
		\caption{VH-MPC with Adaptive Terminal Constraint Sequence (ATCS)}\label{alg:control}
		\begin{algorithmic}[1]
			\State \textbf{Input} $\xx(0)$, $\lambda$
			\State Solve $\mathbb{P}_0(\xx(0),\{0\})$ and get $(J_0^*,\,N_0^*,\,\vv_0^*,\,\zz_0^*)$
			\State {$\ZZ_{f,0} \gets \{0\}$}
			\State $\bar{N} \gets N_0^*$
			\State $\uu(0) \gets \vv^*_0(0)$
			\State $\xx({1}) \gets \AA\xx(0)+\BB\uu(0)+\ww(0)$
			\State $k \gets 0$
			\While{$N_k^*>1$} 
			\State $k \gets k+1$
			\State \textit{(C1)} Solve $\mathbb{P}_k(\xx(k),\{0\})$, get $(\tilde{J}_k^*, \tilde{N}_k^*,\tilde{\vv}_k^*,\tilde{\zz}_k^*)$ 
			\If{$\tilde{J}_k^* > J^*_{k-1} -\lambda$}
			\State $\ZZ_{f,k} \gets \ZZ_{f,k-1} \oplus \AA_K^{N^*_{k-1}-1}\WW$
			\State  \textit{(C2)} Solve ${\mathbb{P}}_k(\xx(k),\ZZ_{f,k})$ with $N_k   \leq   N^*_{k-1}-1$,
			\NoNumber{ get $({J}_k^*,{N}_k^*,{\vv}_k^*,{\zz}_k^*)$ }
			\Else  
			\State $({J}_k^*,{N}_k^*,{\vv}_k^*,{\zz}_k^*) \gets (\tilde{J}_k^*, \tilde{N}_k^*,\tilde{\vv}_k^*,\tilde{\zz}_k^*)$
			\State  $\ZZ_{f,k} \gets \{0\}$
			\State $\bar{N}\gets N_k^*$
			\EndIf
			\State $\uu(k) \gets \vv^*_k(0)$
			\State $\xx({k+1}) \gets \AA\xx(k)+\BB\uu(k)+\ww(k)$
			\EndWhile
			\State \Return $\bar{N}$
		\end{algorithmic}
	\end{algorithm}
	As it is common in MPC, it is assumed that the optimization problem is feasible at the initial time $k=0$.
	\begin{assum}\label{assum:P0}
		The problem $\mathbb{P}_0(\xx(0),\{0\})$ is feasible.
	\end{assum}
	The following result establishes the key properties of the proposed scheme.
	\begin{thm}\label{thm:convergence}
		Let Assumption~\ref{assum:P0} be satisfied. Then, the following statements hold:
		\begin{enumerate}
			\item[(i)] Problems $\mathbb{P}_{k}(\xx(k),\ZZ_{f,k})$, where the sets $\ZZ_{f,k}$ are defined according to Algorithm~\ref{alg:control}, are feasible for all $k=1,2,\ldots$ and any $\ww(k)\in\WW$;
			\item[(ii)] By selecting $\gamma_v$ and $\gamma_z$ such that
			\begin{equation}\label{eq:lambda}
				\bar{\lambda}  =  1-\underset{\ww\in\WW}{\text{sup}}\biggl\{\gamma_z\sum_{j=0}^{\infty} \|\AA_{\KK}^j\ww\|+\gamma_v\sum_{j=0}^{\infty} \|\KK\AA_{\KK}^j\ww\|\biggr\}>0
			\end{equation}
			and setting $\lambda = \bar{\lambda}$ in Algorithm~\eqref{alg:control}, the optimal cost decreases by at least $\bar{\lambda}$ at each step, i.e.
			\begin{equation}\label{eq:cost_decrease}
				J_{k+1}^*\leq J_k^*-\bar{\lambda},\,\forall k.
			\end{equation}
		\end{enumerate}
	\end{thm}
	
	\begin{proof} 
		(i) To prove the statement it suffices to show that if problem $\mathbb{P}_{k}(\xx(k),\ZZ_{f,k})$ is feasible, then also problem $\mathbb{P}_{k+1}(\xx(k+1),\ZZ_{f,k+1})$ is feasible. At step $k$, the solution $\boldsymbol{\vv}_k^*,\,\boldsymbol{\zz}_k^*$ with optimal cost $J_k^*$ can result from either problem $\mathbb{P}_k(\xx(k),\{0\})$ (hereafter denoted as case \textit{(C1)}) or $\mathbb{P}_k(\xx(k),\ZZ_{f,k})$, with $\ZZ_{f,k}$ chosen as in Algorithm~\ref{alg:control} (case \textit{(C2)}). Feasibility of either problem implies that
		\begin{align}
			&\zz_k^*(j)\in\ZZ_k(j) = \XX(k+j)\ominus \SSS(j),\, j = 1,\ldots,N^*_k-1 \label{eq:state1}\\
			&\vv_k^*(j)\in\VV_k(j) = \UU(k+j)\ominus \KK\SSS(j), \, j = 0,\ldots,N^*_k-1 \label{eq:state2}\\
			&\zz_{k}^*(j+1) = \AA\zz_{k}^*(j)+\BB\vv_{k}^*(j), \, j = 0,\ldots,N^*_k-1 \label{eq:state3}\\
			&\zz_k^*(N_k^*) \in \begin{cases}
				\{\rr(k+N_k^*)\} & \text{for case \emph{(C1)}}\\
				\{\rr(k+N_k^*)\} \oplus \ZZ_{f,k} & \text{for case \emph{(C2)}.}
			\end{cases}\label{eq:state4}
		\end{align}
		Consider now the following candidate solution for step $k+1$ with length $N_k^*-1$:
		\begin{equation}\label{eq:zv_candidate}
			\begin{array}{l}
				\hat{\zz}_{k+1}(j) =\zz^*_k(j+1) + \AA_K^j\ww(k), \quad j = 0,\ldots,N^*_k-1\\
				\hat{\vv}_{k+1}(j) = \vv^*_k(j+1) + \KK\AA_K^j\ww(k), \quad j = 0,\ldots,N^*_k-2,
			\end{array}
		\end{equation}
		with associated cost 
		\begin{equation}\label{eq:Jhat}
				\hat{J}_{k+1} = N_k^* -1 +\gamma_z    \sum_{j=0}^{N_k^*-1}   \|\hat{\zz}_{k+1} (j) - \rr(k+j+1)\| + \gamma_v    \sum_{j=0}^{N_k^*-2}   \|\hat{\vv}_{k+1}(j)\| .
		\end{equation}
		
		It turns out that $(\hat{J}_{k+1},\,N_k^*-1,\,\hat{\vv}_{k+1},\hat{\zz}_{k+1})$ is a feasible solution for problem $\mathbb{P}_{k+1}(\xx(k+1),\ZZ_{f,k+1})$. In fact, the initial constraint $\zz_{k+1}(0)=\xx(k+1)$ is trivially satisfied. Moreover, by using~\eqref{eq:state3} and \eqref{eq:zv_candidate}, we get:
		\begin{equation*}
			\begin{split}
				&\hat{\zz}_{k+1}(j+1)-\AA_{\KK}^{j+1}\ww(k)\\ 
				&= \AA\left(\hat{\zz}_{k+1}(j)-\AA_{\KK}^j\ww(k)\right)+\BB\left(\hat{\vv}_{k+1}(j)-\KK\AA_{\KK}^j\ww(k)\right) \\
				&= \AA\hat{\zz}_{k+1}(j)+\BB\hat{\vv}_{k+1}(j) - \AA_{\KK}^{j+1}\ww(k).
			\end{split}
		\end{equation*}
		Hence,  $\hat{\zz}_{k+1}(j+1) = \AA\hat{\zz}_{k+1}(j)+\BB\hat{\vv}_{k+1}(j).$
		Furthermore,  regarding the state constraints, from~\eqref{eq:state1} one has that
		\begin{equation*}
			\begin{split}
				\hat{\zz}_{k+1}(j) &= \zz_k^*(j+1) + \AA_{\KK}^j\ww(k) \in \ZZ_k(j+1)\oplus\AA_{\KK}^j\WW \\
				&= \XX(k+j+1)\ominus \SSS(j+1) \oplus\AA_{\KK}^j\WW \\
				&= \XX(k+j+1)\ominus \sum_{i=0}^{j} \AA_{\KK}^i\WW \oplus\AA_{\KK}^j\WW \\
				&=\XX(k+j+1)\ominus \left[\sum_{i=0}^{j-1} \AA_{\KK}^i\WW \oplus\AA_{\KK}^j\WW \right]  \oplus \AA_{\KK}^j\WW \\
				&\subseteq \XX(k+j+1)\ominus \SSS(j) = \ZZ_{k+1}(j).
			\end{split}
		\end{equation*}	
		Similarly, for the input constraints, using~\eqref{eq:state2} one gets:
		\begin{equation*}
			\begin{split}
				\hat{\vv}_{k+1}(j) &= \vv_k^*(j+1) + \KK\AA_{\KK}^j\ww(k)  \\
				& \in \VV_k(j+1)\oplus\KK\AA_{\KK}^j\WW  \\
				&= \UU(k+j+1)\ominus \KK\SSS(j+1) \oplus\KK\AA_{\KK}^j\WW  \\
				&= \UU(k+j+1)\ominus \KK\sum_{i=0}^{j} \AA_{\KK}^i\WW \oplus \KK\AA_{\KK}^j\WW  \\
				&\subseteq \UU(k+j+1)\ominus \KK\SSS(j) = \VV_{k+1}(j).
			\end{split}
		\end{equation*}		
		As for the terminal constraint, cases\textit{ (C1)} and \textit{(C2)} need to be analyzed separately. In the former, we have that $\zz_{k}(N_k^*) =\rr(k+N_k^*)$ and $\ZZ_{f,k}=\{0\}$, at step $k$. Hence, at step $k+1$, one has $\ZZ_{f,k+1} = \AA_K^{N^*_k-1}\WW$, and thus:
		\begin{equation*}
			\begin{split}
				\hat{\zz}_{k+1}(N_k^* - 1) &= \zz_k^*(N_k^*) + \AA_{\KK}^{N_k^*-1}\ww(k)  \\ 
				&\in \{\rr(k+N_k^*)\}\oplus \AA_{\KK}^{N_k^*-1}\WW \\
				& = \{\rr(k+N_k^*)\}\oplus\ZZ_{f,k+1}.
			\end{split}
		\end{equation*}
		For case \textit{(C2)}, first notice that the candidate solution \eqref{eq:zv_candidate}
		satisfies the length constraint $N_{k+1}   \leq   N^*_{k}-1$. Then, by exploiting~\eqref{eq:state4} and~\eqref{eq:Zfk} one has:
		\begin{equation*}
			\begin{split}
				\hat{\zz}_{k+1}(N_k^* - 1) &= \zz_k^*(N_k^*) + \AA_{\KK}^{N_k^*-1}\ww(k)  \\ 
				&\in \{\rr(k+N_k^*)\}\oplus \ZZ_{f,k} \oplus \AA_{\KK}^{N_k^*-1}\WW\\
				& = \{\rr(k+N_k^*)\}\oplus \ZZ_{f,k+1}.
			\end{split}
		\end{equation*}
		Therefore, problem $\mathbb{P}_{k+1}(\xx(k+1),\ZZ_{f,k+1})$ is feasible for both cases \textit{(C1)} and \textit{(C2)}.
		
		(ii) The cost $\hat{J}_{k+1}$ associated to the candidate solution~\eqref{eq:zv_candidate} satisfies
		\begin{equation*}
			\begin{array}{l}
				\hat{J}_{k+1}  = N_k^*-1 + \gamma_z \sum_{j=0}^{N_k^*-1}\|\hat{\zz}_{k+1}(j)-\rr(k+j+1)\| \\[3mm]
				+ \gamma_v \sum_{j=0}^{N_k^*-2}\|\hat{\vv}_{k+1}(j)\|\\[3mm]
				= N_k^* - 1+  \gamma_z \sum_{j=0}^{N_k^*-1}\|\zz_k^*(j + 1)+\AA_{\KK}^j\ww(k) - \rr(k + j + 1)\| \\[3mm]
				+ \gamma_v \sum_{j=0}^{N_k^*-2}\|\vv_k^*(j+1)+\KK\AA_{\KK}^j\ww(k)\|\\[3mm]
				\leq N_k^*-1+ \gamma_z \sum_{j=0}^{N_k^*-1}\|\zz_k^*(j+1)- \rr(k+j+1)\| \\[3mm] 	
				+\gamma_z\sum_{j=0}^{N_k^*-1}\|\AA_{\KK}^j\ww(k)\|  + \gamma_v \sum_{j=0}^{N_k^*-2}\|\vv_k^*(j+1)\|\\[3mm]
				+\gamma_v \sum_{j=0}^{N_k^*-2}\|\KK\AA_{\KK}^j\ww(k)\|\\[3mm]
				=N_k^*-1+ \gamma_z \sum_{j=0}^{N_k^*-1}\|\zz_k^*(j+1)- \rr(k+j+1)\| \\[3mm] 	
				+\gamma_z\sum_{j=0}^{N_k^*-1}\|\AA_{\KK}^j\ww(k)\|  + \gamma_v \sum_{j=0}^{N_k^*-2}\|\vv_k^*(j+1)\|\\[3mm]
				+\gamma_v \sum_{j=0}^{N_k^*-2}\|\KK\AA_{\KK}^j\ww(k)\|+ \gamma_z\|\zz_k^*(0)- \rr(k)\|\\[3mm]
				- \gamma_z\|\zz_k^*(0)- \rr(k)\| + \gamma_v\|\vv_k^*(0)\| - \gamma_v\|\vv_k^*(0)\| \\[3mm]
				\leq J_k^* -  1 +  \gamma_v  \sum_{j=0}^{N_k^*-2}\|\KK\AA_{\KK}^j\ww(k)\| + \gamma_z \sum_{j=0}^{N_k^*-1}\|\AA_{\KK}^j\ww(k)\| \\[3mm]  \leq J_k^* - \bar{\lambda},
			\end{array}
		\end{equation*}
		where $\bar{\lambda}$ is selected as in~\eqref{eq:lambda}.
		Hence, the optimal cost at time $k+1$ satisfies $$J_{k+1}^* \leq \hat{J}_{k+1}\leq J_k^* - \bar{\lambda}.$$ $\hfill\blacksquare$
	\end{proof}
	The next result states finite-time convergence of the proposed control strategy.
	\begin{thm}\label{thm:Nbar}
		Let Assumption~\ref{assum:P0} be satisfied. Then, the trajectories $\xx(k)$ of the uncertain closed-loop system~\eqref{eq:sys}-\eqref{eq:u_RMPC}, with the VH-MPC control law defined via Algorithm~\ref{alg:control}, converge in a finite number of steps $N_{ct}$ to the set $\{\rr(N_{ct})\}\oplus\SSS(\bar{N})$, being $\bar{N}$ as returned by Algorithm~\ref{alg:control}. Moreover, $N_{ct} \leq \lfloor J_0^*/\bar{\lambda} \rfloor$ where $J_0^*$ is the optimal cost of the initial problem $\mathbb{P}_0(\xx(0),\{0\})$.
	\end{thm}
	\begin{proof} 
		Finite-time convergence in $N_{ct} \leq \lfloor J_0^*/\bar{\lambda} \rfloor$ steps is a direct consequence of the decreasing property of the cost in~\eqref{eq:cost_decrease} ensured by Theorem~\ref{thm:convergence}. 
		Now, let $\bar{k}$ be the algorithm iteration at which line 17 of Algorithm ~\ref{alg:control} is executed for the last time ($\bar{k}=0$ if it is never executed).
		Due to the adaptive choice of the terminal set in Algorithm~\ref{alg:control}, the final constraint of the optimization problem solved at the last step $k=N_{ct}-1$ is $\zz_k(1) \in \{\rr(N_{ct})\}\oplus \ZZ_{f,N_{ct}-1}$, where $\ZZ_{f,N_{ct}-1} = \sum_{i=\bar{k}}^{N_{ct}-1}\AA_{\KK}^{N_i^*-1} \WW$ if $\bar{k}<N_{ct}$, and $\zz_k(1) = \rr(N_{ct})$ if $\bar{k}=N_{ct}$. Therefore, the final state of the closed-loop system at completion time $N_{ct}$ satisfies
		\begin{equation*}
			\begin{split}
				\xx(N_{ct})& =\AA\xx(N_{ct}-1)+\BB\uu(N_{ct}-1)+\ww(N_{ct}-1) \\ &=\zz_k(1)+\ww(N_{ct}-1) \\ 				
				&\in \{\rr(N_{ct})\}\oplus \sum_{i=\bar{k}}^{N_{ct}-1}\AA_{\KK}^{N_i^*-1} \WW \oplus \WW \\
				& \subseteq \{\rr(N_{ct})\} \oplus \SSS(\bar{N}).
			\end{split}
		\end{equation*}
		where the last  inclusion stems from $N_{\bar{k}}^*=\bar{N}$ and the fact that the sequence $N_i^*$ for $\bar{k}\leq i \leq N_{ct}-1$, is strictly decreasing by construction, due to the constraint $N_k   \leq   N^*_{k-1}-1$ in (C2) (see line 13 of Algorithm ~\ref{alg:control}).		$\hfill\blacksquare$
	\end{proof}
	
	\begin{rem}\label{remark:small_set}
		The worst-case outcome of Algorithm~\ref{alg:control} is $\bar{N} = N_{0}^*$. In this case, the final state lies in the set $\{\rr(N_{ct})\}\oplus \SSS(N_{0}^*)$. Note that the latter is always smaller than $\SSS(\infty)$, which is the performance guaranteed by the approach borrowed from  \cite{richards2006robust}. However, in most cases, the condition in line 11 of Algorithm~\ref{alg:control} is triggered only in close proximity of the target, i.e., for small values of $N_{k-1}^*$. Hence, $\bar{N}$ is typically significantly smaller than $N_{0}^*$. In turn, this leads to $\SSS({\bar{N}})$ being much smaller than $\SSS(\infty)$, as observed in the case studies worked out (see Section~\ref{sec:simulations}).
	\end{rem}
	\begin{rem}\label{remark:computational}
		Problem~\eqref{eq:mpc_tube} is a convex program with integrality constraints, which typically involves computational challenges. When the 1-norm is used in cost~\eqref{eq:cost}, the problem becomes a Mixed-Integer Linear Program (MILP). The upper bound on the completion time stated in Theorem~\ref{thm:Nbar} allows one to restrict the set of admissible horizon lengths, thus reducing the computational burden. Moreover, it is possible to remarkably speed up computations, by exploiting recursive methods such as that proposed in \cite{persson2024optimization}.
	\end{rem}
	
	\subsection{Robust Minimum-Time MPC}
	A special instance of the VH-MPC problem is minimum-time MPC in which one has $\gamma_z = \gamma_v=0$ and the cost function in~\eqref{eq:cost} boils down to $J = N$. In this case, the analysis of recursive feasibility and finite-time convergence is more straightforward, as clarified in the next result. 
	\begin{cor}\label{cor:convergence_MT}
		Let Assumption~\ref{assum:P0} be satisfied. If $\,\gamma_z=\gamma_v=0$, then problems $\mathbb{P}_k(\xx(k),\ZZ_{f,k})$, with $\ZZ_{f,k}$ selected according to Algorithm~\ref{alg:control}, are feasible for all $k = 1,2,\ldots$ and the optimal horizon length decreases by at least 1 at each step $k$.
	\end{cor}
	\begin{proof} 
		Recursive feasibility is inherited from the proof of Theorem~\ref{thm:convergence}. With $\gamma_z=\gamma_v=0$, the quantity $\bar{\lambda}$ in~\eqref{eq:lambda} turns out to be 1, indicating that the condition to be checked in line 11 of Algorithm~\ref{alg:control} becomes $\tilde{N}_k^* > N_{k-1}^* - 1$. If such a condition holds, problem $\mathbb{P}_k(\xx(k),\ZZ_{f,k})$ is solved. According to the proof of Theorem~\ref{thm:convergence}, a solution of length $N_{k-1}^*-1$ is feasible. Hence, being $J_k=N_k$, one gets $N_k^*\leq N_{k-1}^*-1$. Conversely, if $\tilde{N}_k^*\leq N_{k-1}^*-1$, Algorithm~\ref{alg:control} sets $N_k^*=\tilde{N}_k^*$. Consequently, the proposed control strategy ensures that the optimal horizon length decreases by at least 1 at every time step $k\geq 1$.	 $\hfill\blacksquare$ 
	\end{proof}
	As a direct consequence of Corollary~\ref{cor:convergence_MT}, the finite-time convergence property claimed in Theorem~\ref{thm:Nbar} reads as follows for the minimum-time MPC problem.
	\begin{cor}\label{cor:Nbar_MT}
		Let Assumption~\ref{assum:P0} be satisfied and $\gamma_z=\gamma_v=0$. Then, finite-time convergence of $\xx(k)$ to  $\{r(N_{ct})\}\oplus\SSS(\bar{N})$ is achieved in $N_{ct} \leq N_0^*$ steps, being $N_0^*$ the solution of the initial problem $\mathbb{P}_0(\xx(0),\{0\})$.
	\end{cor}
	\begin{rem}
		In minimum-time MPC, the optimal horizon length is ensured to decrease over time, as stated in Corollary~\ref{cor:convergence_MT}. Consequently, at each step $k$ of Algorithm~\ref{alg:control}, the optimization variable $N_k$ can be effectively upper-bounded by $N_{k-1}^* - 1$. This information can be used to further speed up the solution of problem~\eqref{eq:mpc_tube}.
	\end{rem}
	
	\section{Numerical Results}\label{sec:simulations}
	In this section, the proposed control algorithm is tested on two case studies. In the first, a double integrator is considered to illustrate the main features of the proposed VH-MPC strategy with adaptive terminal constraint sequence (in the following, denoted by ATCS). The second case study considers a realistic scenario, in which a controlled satellite is required to intercept an uncontrolled tumbling object (such as a defunct satellite or space debris). The performance of ATCS is compared to that of the VH-MPC control law with fixed terminal constraint sequence (denoted by FTCS) borrowed from~\cite{richards2006robust}.
	
	In both examples, the 1-norm is employed in the cost function $J$ in~\eqref{eq:cost}. The motivation of this choice is twofold. Firstly, formulating problem~\eqref{eq:mpc_tube} with the 1-norm enables it to be treated as a mixed-integer-linear-program (MILP), which features a lighter computational burden compared to other mixed integer optimization problems. Secondly, the 1-norm explicitly accounts for fuel optimization in the considered aerospace application (see, e.g., \cite{leomanni2019sum}).
	
	\subsection{Double Integrator}\label{sec:DI}
	
	Consider the discrete-time double integrator with additive disturbance described by~\eqref{eq:sys}, where $\xx(k) = \left[\xx_1(k)\,\,\,\xx_2(k)\right]^T$ and
	\begin{equation*}\label{eq:AB_DI}
		\AA=
		\left[
		\begin{array}{c c}
			1&\quad 1\\
			0&\quad 1 \end{array}\right]
		\quad
		\BB=\left[
		\begin{array}{lll}
			0\\
			1\end{array}\right].
	\end{equation*}
	The control objective is to drive the state to the origin, i.e., $\rr(k) = 0,\,\forall k$, counteracting the presence of the bounded disturbance $\ww(k) \in \WW = \left[-0.1,\,0.1\right] \times \left[-0.4,\,0.4\right]$. Moreover, state and input constraints must be satisfied, specifically: $-25\leq\xx_1(k)\leq 25$, $-2\leq\xx_2(k)\leq 2$ and $|\uu(k)|\leq 2$. The control gain in~\eqref{eq:u_RMPC} is set as $\KK=\left[-0.06\,\,\,-0.5\right]$ so that $\AA_{\KK}$ has eigenvalues 0.7 and 0.8. The weights in the cost function are $\gamma_z = 0.02$ and $\gamma_v = 1$, corresponding to $\bar{\lambda} = 0.27$ in~\eqref{eq:lambda}.

	\begin{figure}[t]
		\centering
		\includegraphics[width=0.7\columnwidth]{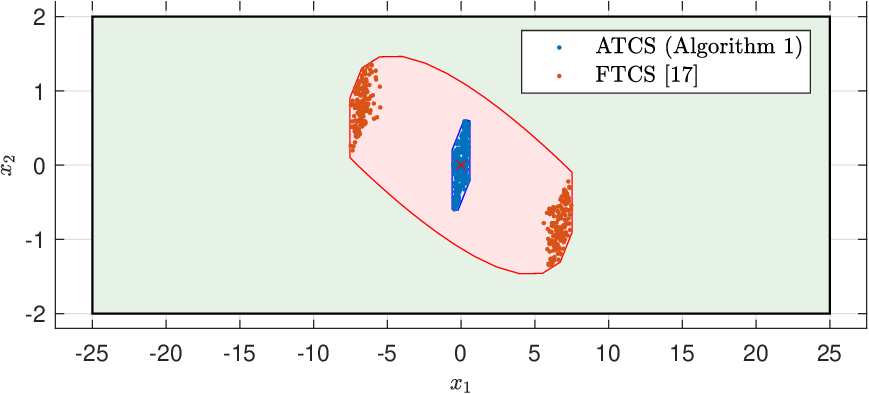}
		\includegraphics[width=0.7\columnwidth]{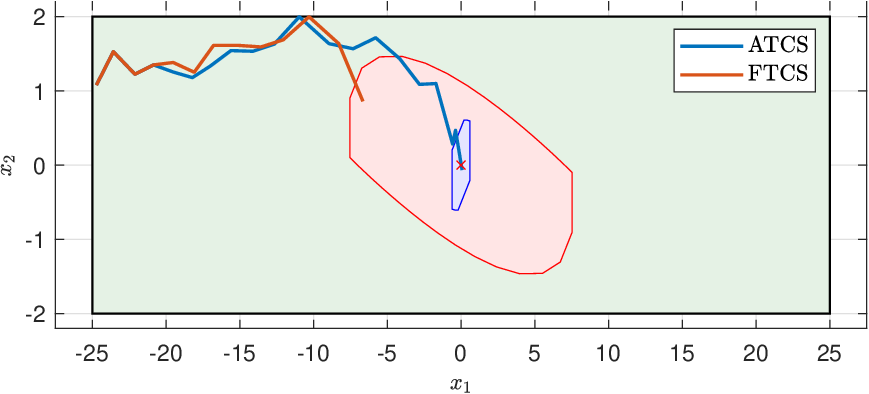}
		\caption{Top: final states obtained by applying ATCS (blue dots) and FTCS (red dots). Bottom: one of the simulated trajectories. In both panes: the red set is $\SSS(\infty)$; the blue set is $\SSS(\bar{N})$, with $\bar{N}=2$. The green rectangle is $\XX$.}\label{fig:DI_MC_final}
	\end{figure}

	A Monte Carlo simulation across 300 different scenarios has been conducted. For each scenario, the initial condition $\xx(0)$ has been uniformly sampled over the feasible region (i.e., the values of $\xx(0)$ for which the problem $\mathbb{P}_0(\xx(0),\{0\})$ is feasible, as required by Assumption~\ref{assum:P0}), excluding elements starting inside the set $\SSS(\infty)$, for which the FTCS approach is meaningless. The disturbance sequences have been sampled uniformly from the set $\WW$ for each simulated scenario. Figure~\ref{fig:DI_MC_final} displays the final states for both control laws. It can be observed that ACTS drives the states to a relatively small neighborhood of the origin, achieving a mean terminal distance of 0.38. Conversely, the FTCS control strategy drives the state trajectories much farther away from the origin, with an average final state norm of 6.74. Furthermore, in the latter case, it can be noticed that final states generally lie near the boundary of $\SSS(\infty)$. More specifically, as noticeable in~\eqref{eq:Zf_How}, smaller values of $N_k$ (implying a small $\SSS(N_k)$) result in larger sets $\ZZ_{f,k}$. On the other side, the most frequent $\bar{N}$ returned by ACTS is 2. Thus the majority of the blue points in figure~\ref{fig:DI_MC_final} fall in the set $\SSS(2)$ (depicted in blue). The largest $\bar{N}$ returned by Algorithm~\ref{alg:control} in the 300 tests is 3, thus the adaptive mechanism for the terminal constraint set is triggered only in close proximity of the target (see Remark~\ref{remark:small_set}).  
	In the simulations worked out, the average completion time $N_{ct}$ turns out to be 13 steps, while the average value of $\lfloor J_0^*/\bar{\lambda} \rfloor$ is 73. This is not surprising, as item (ii) in Theorem~\ref{thm:Nbar} provides only a sufficient condition for finite-time convergence.

\begin{figure}[t]
		\centering
		\includegraphics[width=0.7\columnwidth]{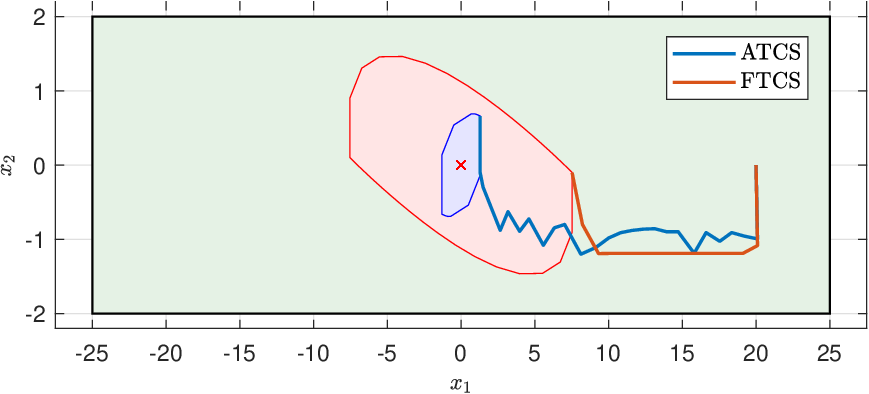}
		\caption{Trajectories resulting from one run of ATCS and FTCS under a persistent disturbance on the boundary of $\WW$. The red and blue sets are $\SSS(\infty)$ and $\SSS(3)$, respectively.}\label{fig:DI_WC_trajectory}
	\end{figure}

	To further demonstrate the performance of the proposed control strategy, a single run is performed with initial condition $\xx(0) = \left[20\,\,\,0\right]^T$ and by assuming a persistent disturbance on the boundary of the set $\WW$, specifically $\ww(k) = \left[0.1\,\,\,0.4\right]^T,\,\forall k$. In Figure~\ref{fig:DI_WC_trajectory}, the trajectory obtained by applying the ATCS strategy (blue) is compared to the one produced by the FTCS strategy (red). It is evident that ATCS achieves a smaller final distance, while the trajectory obtained by applying the FTCS terminates on the boundary of $\SSS(\infty)$ (highlighted in red). In this test, the value of $\bar{N}$ returned by Algorithm~\ref{alg:control} is equal to 3. The final state norm returned by ATCS and FTCS are 1.45 and 7.53, respectively. 
	

	\subsection{Rendezvous with a Tumbling Satellite}\label{sec:RVD}
	In this case study, we consider an orbital rendezvous between a controlled satellite and an uncontrolled tumbling object. This scenario is significant in various contexts, including active debris removal \cite{shan2016review,leomanni2020orbit}.
	
	The dynamics of the servicing satellite are described by the Hill-Clohessy-Wiltshire (HCW) equations \cite{clohessy1960terminal}, written in the radial-transverse-normal (RTN) body frame centered at the target center of mass. The HCW equations are normalized and discretized with sampling interval $\theta_s$ (expressed in radians), leading to the dynamic model
	\begin{equation}\label{zetacirctd}
		\xx(k+1)=\AA \xx(k)+ \BB {\uu}(k)+\ww(k) 
		=e^{ \AA_c  \theta_s\,}  \xx(k)  + \left(\displaystyle\int_{0}^{\theta_s}e^{ \AA_c \theta}\text{d}\theta\right) \BB_c\, \uu(k) + \ww(k),
	\end{equation}
	where $\xx(k) = \left[\xx_p^T(k)\,\,\,\xx_v^T(k)\right]^T $ is the state vector, with $\xx_p(k)$ and $\xx_v(k)$ denoting the position and velocity components; $\uu(k)$ is the acceleration vector along the three axes and
	\begin{equation*}\label{zetacircmattd}
		\AA_c=
		\begin{bmatrix}
			0\,&0\,&0\,&1\,&0\,&0\\
			0\,&0\,&0\,&0\,&1\,&0\\
			0\,&0\,&0\,&0\,&0\,&1\\
			3\,&0\,&0\,&0\,&2\,&0\\
			0\,&0\,&0\,&-2\,&0\,&0\\
			0\,&0\,&-1\,&0\,&0\,&0
		\end{bmatrix}
		\quad
		\BB_c=
		\begin{bmatrix}
			0\,&0\, &0\\
			0\,& 0\, & 0\\
			0\,&  0\,&   0 \\
			1\,& 0\, &    0 \\
			0\,& 1\,& 0  \\
			0\,& 0\, &1
		\end{bmatrix}.
	\end{equation*}
\begin{figure}[t]
\centering
\includegraphics[width=0.5\columnwidth]{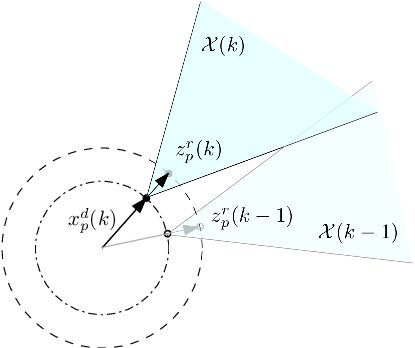}
\caption{Schematic representation of the considered case study in the orbital plane. The dashed line represents the position component of the reference trajectory $\zz_p^r(k)$. The shaded region is the state constraint set.}\label{fig:conf}
\end{figure}
	The state components $\xx_p(k)$ and $\xx_v(k)$ are normalized by $u_{max}/\eta^2$ and $u_{max}/\eta$, respectively, with $\eta$ being the orbit mean motion and $u_{max}$ the maximum deliverable acceleration along each axis. 
	Similarly, the acceleration $\uu(k)$ is normalized by $u_{max}$. Hence, all signals $\xx_p(k)$, $\xx_v(k)$,  $\uu(k)$ and $\ww(k)$ are dimensionless.
	The input constraint set is $\UU(k) = \{\uu:\,\|\uu\|_{\infty} \leq 1\},\,\forall k.$
	The process disturbance $\ww(k)$ is assumed to belong to a box $\WW = \left[-\bar{\ww}_p,\,\bar{\ww}_p\right]^3\times \left[-\bar{\ww}_v,\,\bar{\ww}_v\right]^3 \subset \mathbb{R}^6$ with $\bar{\ww}_p,\,\bar{\ww}_v$ being the maximum disturbances acting on the position and velocity components, respectively.
	The reference nominal trajectory of the target $\rr(k)=\left[\zz_p^r(k)^T\,\,\zz_v^r(k)^T\right]^T$ is represented by the capture point trajectory, where $\zz_p^r(k)$ and $\zz_v^r(k)$ specify the position and velocity components, respectively. The reference trajectory is generated by integrating the classical rotational kinematics of a rigid body, defined by
	\begin{equation}
		\dot{\zz}_p^r(\theta) = \zz_v^r(\theta),\quad \dot{\zz}_v^r(\theta) = \omega(\theta)\times\zz_p^r(\theta),
	\end{equation}
	with $\omega(\theta)$ being the angular velocity vector, and then sampling them with sampling interval $\theta_s$. The capture point rigidly evolves together with the docking port of the target satellite, represented by the vector $\xx^d(k)=\left[\xx_p^d(k)^T\,\,\,\xx_v^d(k)^T\right]^T$.
	The servicing satellite is required to remain inside a visibility cone stemming from the docking port. In this paper, we consider a polytopic inner-approximation of the cone, so that the state constraints $\XX(k)$ are convex polytopes. In particular, the state constraint set is defined as:
	\begin{equation}\label{eq:cone}
		\XX(k) = \Bigl\{\xx_p\in\mathbb{R}^3\!:\!\,\, \|T(k)\left(\xx_p\!-\!\left[\xx_p^T \vec{\xx}_p^d(k)\right]\vec{\xx}_p^d(k)\right)\!\|_\infty  \!\leq\!\dfrac{\tan\alpha}{\sqrt{2}}\!\!\left[\xx_p-{\xx}_p^d(k)\right]^T\!\vec{\xx}_p^d(k) \Bigr\},
	\end{equation}
	where the matrix $T(k)$ defines the planes of the polytopic approximation, $\alpha$ is the half-angle of the visibility cone and $\vec{\xx}_p^d(k)$ is the unit vector indicating the direction of the docking port position at time $k$. 
	Figure~\ref{fig:conf} depicts a schematic representation of the considered setting in the orbital plane. For more details, the reader is referred to \cite{leomanni2022variable}.
	
	The considered rendezvous scenario is set as follows. The target satellite lies on a circular orbit at a distance of 800 km from the Earth, exhibiting tumbling behavior with a spin period of 500 seconds. The spin axis is perpendicular to the orbital plane. Initially, the controlled satellite is positioned 40 m before and 10 m above the target, corresponding to the initial state $\xx(0)=10^{-3}\cdot\left[-2.1857\,\,\,0.5464\,\,\,0\,\,\,0\,\,\,0\,\,\,0\right]^T$. The docking port is located 1.5 meters from the target center of mass, while the capture point is situated at a distance of 1.7 meters. The maximum acceleration of the propulsion system is $u_{max} = 20$ mm/s$^2$, and the sampling interval is $\theta_s = 0.0123$ radians. The visibility cone constraint~\eqref{eq:cone} is characterized by a half-angle $\alpha=\pi/6$ rad. Disturbance bounds are set to $\bar{\ww}_p = 10^{-6}$ and $\bar{\ww}=5\cdot10^{-4}$. The matrix $\KK$ in~\eqref{eq:u_RMPC} is selected such that the poles of $\AA_{\KK}$ are $\{0.6,0.6,0.6,0.5,0.5,0.5\}$. The input and state weights in cost function~\eqref{eq:cost} are set to $\gamma_z=100$ and $\gamma_v=1$, resulting in $\bar{\lambda} = 0.58$ in~\eqref{eq:lambda}.
		\begin{figure}[t]
		\centering
		\includegraphics[width=0.7\columnwidth]{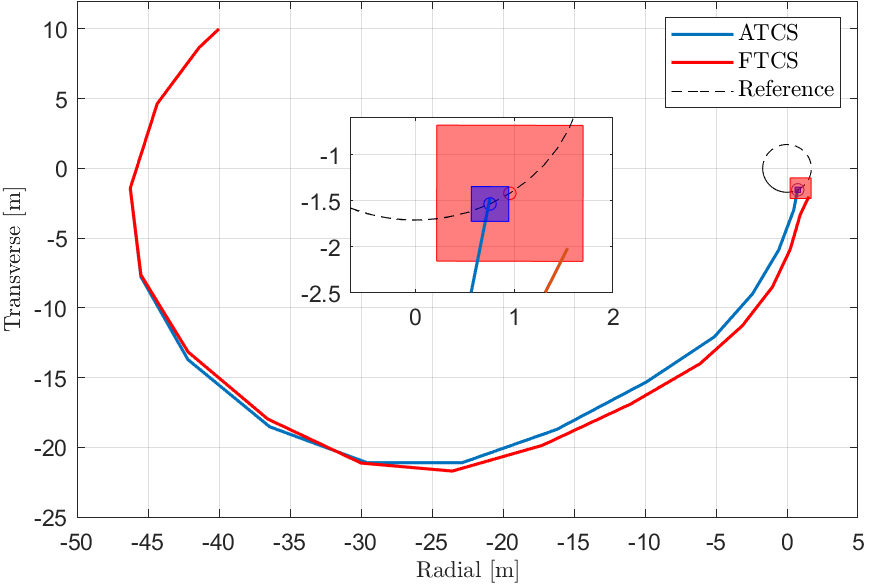}
		\caption{Trajectories resulting from Algorithm~\ref{alg:control} (ATCS) and FTCS. The red and blue sets are $\SSS(\infty)$ and $\SSS(2)$, respectively. The reference trajectory of the capture point is depicted in black.}\label{fig:RVD_trajectory}
	\end{figure}
	\begin{figure}[t]
		\centering
		\includegraphics[width=0.7\columnwidth]{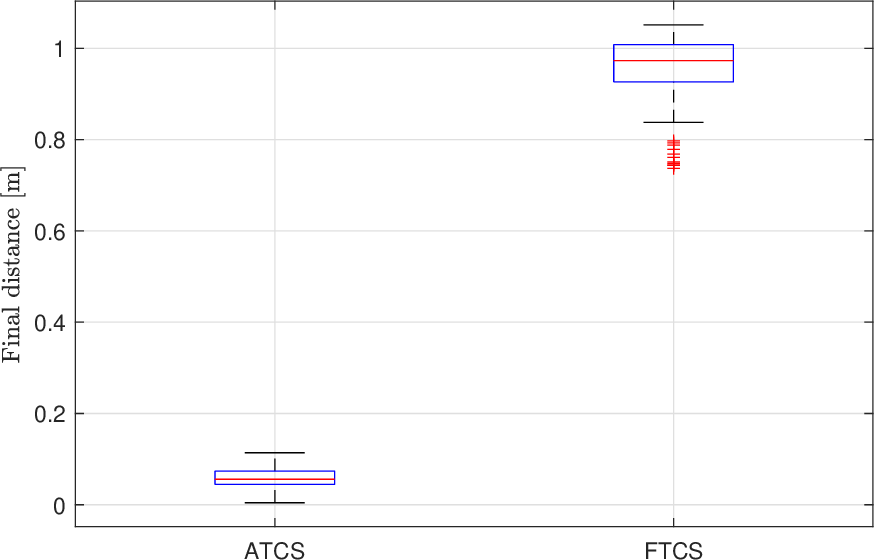}
		\caption{Box plot statistic on the final distance to the target achieved by ATCS and FTCS starting from 100 different initial conditions.}\label{fig:RVD_boxplot}
	\end{figure}

	The proposed control strategy is compared to FTCS. The results are shown in Figure~\ref{fig:RVD_trajectory}, where the radial-transverse trajectories obtained by the two methods are reported. ATCS achieves a much smaller final distance to the selected point on the reference trajectory, equal to 6 cm, as opposed to 88 cm achieved by FTCS.
	Moreover, a Monte Carlo simulation has been conducted across 100 different initial conditions for which Assumption~\ref{assum:P0} holds. A statistic on the final distances achieved by the two compared control laws is reported in Figure~\ref{fig:RVD_boxplot}. It can be seen that the median values of the final distance for ATCS and FTCS are 97~cm and 6~cm, respectively. Notice that the minimum distance reached by FTCS is 74~cm, while the maximum one achieved by ATCS is 11~cm.
	
	
	\section{Conclusions}\label{sec:conclusions}
	A new robust variable-horizon model predictive control scheme has been proposed for discrete-time linear systems
	affected by bounded process disturbances.
	The considered setting encompasses minimum-time MPC as a special case.
	It has been shown that adapting online the terminal constraint set of the MPC optimization problem is useful in
	applications in which it is required to intercept a reference trajectory.
	The proposed solution allows one to remarkably reduce the conservatism of the standard tube-based approach.
	Future investigations will concern the derivation of less conservative bounds on the convergence time of the proposed
	procedure. The extension of the adaptive terminal constraint approach to other variants of robust MPC will be also
	addressed.
	
	\bibliographystyle{ieeetr}        
	\bibliography{aerospace}           

\begin{thebibliography}{10}

\bibitem{rawlings2017model}
J.~B. Rawlings, D.~Q. Mayne, and M.~Diehl, {\em Model predictive control:
  theory, computation, and design}, vol.~2.
\newblock Nob Hill Publishing Madison, WI, 2017.

\bibitem{boyarko2011optimal}
G.~Boyarko, O.~Yakimenko, and M.~Romano, ``Optimal rendezvous trajectories of a
  controlled spacecraft and a tumbling object,'' {\em Journal of Guidance,
  Control, and dynamics}, vol.~34, no.~4, pp.~1239--1252, 2011.

\bibitem{di2012model}
S.~Di~Cairano, H.~Park, and I.~Kolmanovsky, ``Model predictive control approach
  for guidance of spacecraft rendezvous and proximity maneuvering,'' {\em
  International Journal of Robust and Nonlinear Control}, vol.~22, no.~12,
  pp.~1398--1427, 2012.

\bibitem{capello2018flyable}
E.~Capello, F.~Dabbene, G.~Guglieri, and E.~Punta, ```{F}lyable' guidance and
  control algorithms for orbital rendezvous maneuver,'' {\em SICE Journal of
  Control, Measurement, and System Integration}, vol.~11, no.~1, pp.~14--24,
  2018.

\bibitem{leomanni2022variable}
M.~Leomanni, R.~Quartullo, G.~Bianchini, A.~Garulli, and A.~Giannitrapani,
  ``Variable-horizon guidance for autonomous rendezvous and docking to a
  tumbling target,'' {\em Journal of Guidance, Control, and Dynamics}, vol.~45,
  no.~5, pp.~846--858, 2022.

\bibitem{ramirez2023collision}
J.~Ram{\'\i}rez, L.~Felicetti, and D.~Varagnolo, ``Collision-avoiding model
  predictive rendezvous strategy to tumbling launcher stages,'' {\em Journal of
  Guidance, Control, and Dynamics}, vol.~46, no.~8, pp.~1564--1579, 2023.

\bibitem{verscheure2009time}
D.~Verscheure, B.~Demeulenaere, J.~Swevers, J.~De~Schutter, and M.~Diehl,
  ``Time-optimal path tracking for robots: A convex optimization approach,''
  {\em IEEE Transactions on Automatic Control}, vol.~54, no.~10,
  pp.~2318--2327, 2009.

\bibitem{nascimento2018nonholonomic}
T.~P. Nascimento, C.~E. D{\'o}rea, and L.~M.~G. Gon{\c{c}}alves, ``Nonholonomic
  mobile robots' trajectory tracking model predictive control: a survey,'' {\em
  Robotica}, vol.~36, no.~5, pp.~676--696, 2018.

\bibitem{ardakani2018model}
M.~M.~G. Ardakani, B.~Olofsson, A.~Robertsson, and R.~Johansson, ``Model
  predictive control for real-time point-to-point trajectory generation,'' {\em
  IEEE Transactions on Automation Science and Engineering}, vol.~16, no.~2,
  pp.~972--983, 2018.

\bibitem{van2011model}
L.~Van~den Broeck, M.~Diehl, and J.~Swevers, ``A model predictive control
  approach for time optimal point-to-point motion control,'' {\em
  Mechatronics}, vol.~21, no.~7, pp.~1203--1212, 2011.

\bibitem{rosmann2015timed}
C.~R{\"o}smann, F.~Hoffmann, and T.~Bertram, ``Timed-elastic-bands for
  time-optimal point-to-point nonlinear model predictive control,'' in {\em
  2015 european control conference (ECC)}, pp.~3352--3357, IEEE, 2015.

\bibitem{verschueren2017stabilizing}
R.~Verschueren, H.~J. Ferreau, A.~Zanarini, M.~Mercang{\"o}z, and M.~Diehl, ``A
  stabilizing nonlinear model predictive control scheme for time-optimal
  point-to-point motions,'' in {\em 2017 IEEE 56th annual conference on
  decision and control (CDC)}, pp.~2525--2530, IEEE, 2017.

\bibitem{krener2018adaptive}
A.~J. Krener, ``Adaptive horizon model predictive control,'' {\em
  IFAC-PapersOnLine}, vol.~51, no.~13, pp.~31--36, 2018.

\bibitem{sutherland2019closed}
R.~L. Sutherland, I.~V. Kolmanovsky, A.~R. Girard, F.~A. Leve, and C.~D.
  Petersen, ``On closed-loop {L}yapunov stability with minimum-time {MPC}
  feedback laws for discrete-time systems,'' in {\em 2019 IEEE 58th Conference
  on Decision and Control (CDC)}, pp.~5231--5237, IEEE, 2019.

\bibitem{greer2020shrinking}
W.~B. Greer and C.~Sultan, ``Shrinking horizon model predictive control method
  for helicopter--ship touchdown,'' {\em Journal of Guidance, Control, and
  Dynamics}, vol.~43, no.~5, pp.~884--900, 2020.

\bibitem{farooqi2020shrinking}
H.~Farooqi, L.~Fagiano, P.~Colaneri, and D.~Barlini, ``Shrinking horizon
  parametrized predictive control with application to energy-efficient train
  operation,'' {\em Automatica}, vol.~112, p.~108635, 2020.

\bibitem{richards2006robust}
A.~Richards and J.~P. How, ``Robust variable horizon model predictive control
  for vehicle maneuvering,'' {\em International Journal of Robust and Nonlinear
  Control}, vol.~16, no.~7, pp.~333--351, 2006.

\bibitem{wang2021path}
H.~Wang, Q.~Wang, W.~Chen, L.~Zhao, and D.~Tan, ``Path tracking based on model
  predictive control with variable predictive horizon,'' {\em Transactions of
  the Institute of Measurement and Control}, vol.~43, no.~12, pp.~2676--2688,
  2021.

\bibitem{ngo2022variable}
T.~D. Ngo and C.~Sultan, ``Variable horizon model predictive control for
  helicopter landing on moving decks,'' {\em Journal of Guidance, Control, and
  Dynamics}, vol.~45, no.~4, pp.~774--780, 2022.

\bibitem{persson2024optimization}
L.~Persson, A.~Hansson, and B.~Wahlberg, ``An optimization algorithm based on
  forward recursion with applications to variable horizon {MPC},'' {\em
  European Journal of Control}, vol.~75, p.~100900, 2024.

\bibitem{chisci2001systems}
L.~Chisci, J.~A. Rossiter, and G.~Zappa, ``Systems with persistent
  disturbances: predictive control with restricted constraints,'' {\em
  Automatica}, vol.~37, no.~7, pp.~1019--1028, 2001.

\bibitem{kolmanovsky1998theory}
I.~Kolmanovsky and E.~G. Gilbert, ``Theory and computation of disturbance
  invariant sets for discrete-time linear systems,'' {\em Mathematical problems
  in engineering}, vol.~4, pp.~317--367, 1998.

\bibitem{leomanni2019sum}
M.~{Leomanni}, G.~{Bianchini}, A.~{Garulli}, A.~{Giannitrapani}, and
  R.~{Quartullo}, ``Sum-of-norms model predictive control for spacecraft
  maneuvering,'' {\em IEEE Control Systems Letters}, vol.~3, no.~3,
  pp.~649--654, 2019.

\bibitem{shan2016review}
M.~Shan, J.~Guo, and E.~Gill, ``Review and comparison of active space debris
  capturing and removal methods,'' {\em Progress in Aerospace Sciences},
  vol.~80, pp.~18--32, 2016.

\bibitem{leomanni2020orbit}
M.~Leomanni, G.~Bianchini, A.~Garulli, A.~Giannitrapani, and R.~Quartullo,
  ``Orbit control techniques for space debris removal missions using electric
  propulsion,'' {\em Journal of Guidance, Control, and Dynamics}, vol.~43,
  no.~7, pp.~1259--1268, 2020.

\bibitem{clohessy1960terminal}
W.~Clohessy and R.~Wiltshire, ``Terminal guidance system for satellite
  rendezvous,'' {\em Journal of the aerospace sciences}, vol.~27, no.~9,
  pp.~653--658, 1960.

\end{thebibliography}

	
\end{document}